\documentclass{webofc}

\usepackage[varg]{txfonts}   
\usepackage{hyperref}
\usepackage{url}
\hypersetup{colorlinks=true,citecolor=blue,urlcolor=blue,linkcolor=blue}
\usepackage{amsmath}
\usepackage{amsfonts}	
\usepackage{amssymb}
\usepackage{nccmath}    
\usepackage{multirow}
\usepackage{booktabs}   
\usepackage[dvipsnames]{xcolor}

\usepackage{mathtools}  
\usepackage{bbold}		

\newcommand{\diff}{{\rm d}} 
\newcommand{\im}{{\rm i}} 

\newcommand{\bm}[1]{\boldsymbol{#1}}
\newcommand{\bvec}[1]{\boldsymbol{#1}}

\newcommand{\kappabar}{{\bar{\kappa}}}
\newcommand{\dn}{\delta\hspace{-1pt}n}

\newcommand\mydef{\mathrel{\stackrel{\makebox[0pt]{\mbox{\normalfont\tiny{def}}}}{=}}}

\begin{document}
%
%
\title{Gluon emission by a $\text{q}\overline{\text{q}}$ antenna with realistic parton-medium interactions}
%
%

\author{
\firstname{Carlota}         \lastname{Andres}\inst{1,2}       
\and \firstname{Liliana}    \lastname{Apolinário}\inst{2,3}   
\and \firstname{Néstor}     \lastname{Armesto}\inst{4}    
\and \firstname{André}      \lastname{Cordeiro}\inst{2,3}     
\fnsep\thanks{\email{andre.cordeiro@tecnico.ulisboa.pt}}
\and \firstname{Fabio}      \lastname{Dominguez}\inst{4}    
\and \firstname{Pablo}      \lastname{Guerrero-Rodríguez}\inst{4}     
\and \firstname{José Guilherme}  \lastname{Milhano}\inst{2,3}      
}

\institute{%
Center for Theoretical Physics – a Leinweber Institute, Massachusetts Institute of Technology,
\\ 
Cambridge, MA 02139, USA 
\and
Laboratório de Instrumentação e Física Experimental de Partículas, 
\\ 
Av. Prof. Gama Pinto, 2, 1649-003, Lisbon, Portugal
\and
Departamento de Física, Instituto Superior Técnico, Universidade de Lisboa, 
\\ 
Av. Rovisco Pais 1, 1049-001 Lisbon, Portugal
\and 
Instituto Galego de F\'{\i}sica de Altas Enerx\'{\i}as (IGFAE), Universidade de Santiago de Compostela,
\\ 
15782 Santiago de Compostela, Galicia-Spain
}

\abstract{The spectrum of coherent gluon radiation from a quark-anti-quark pair experiencing multiple scatterings within a coloured medium is central for understanding in-medium parton cascades. Despite its foundational importance, current results are limited by reliance on simplified scattering rates, such as the harmonic oscillator approximation, valid only in restricted phase-space regions. Using the formalism introduced in \cite{Andres:2020vxs}, we express the gluon emission spectrum as a set of differential equations that can be solved numerically, circumventing conventional approximations. We present the transverse momentum and energy distributions of emitted gluons for realistic interaction models, illustrating the breakdown of colour coherence across the entire accessible phase-space, and consequently enabling a higher-precision description of jet observables.}
\maketitle
%

\section{Introduction and Derivation}

Recent progress on jet quenching phenomenology \cite{Apolinario:2024equ} has revealed the versatility of jet observables as tools to study different stages of heavy ion collisions \cite{Andres:2019eus,Apolinario:2020uvt,Apolinario:2024hsm}, and highlighted the importance of precision calculations of in-medium radiative processes. 
In this work, we revisit the emission of soft medium-induced gluons with light-cone and transverse momentum $(\omega, \bm{k})$ from an antenna consisting of a quark and anti-quark with momenta $(p^+,\bm{p})$ and $(\bar{p}^+,\bm{\bar{p}})$ respectively.
The emission spectrum, first calculated in \cite{Mehtar-Tani:2011lic,Casalderrey-Solana:2011ule}, can be expressed in terms of the gluon transverse momenta relative to the quark (anti-quark) $\bm{\kappa} = \bm{k} - \frac{\omega}{p^+} \bm{p}$ ($\bm{\kappabar} = \bm{k} - \frac{\omega}{\bar{p}^+} \bm{\bar{p}}$):
\begin{equation}
    \omega\dfrac{\diff^3 I}{\diff\omega \diff^2\bvec{k}} 
    = 
    \dfrac{\alpha_s}{(2\pi)^2 \omega^2} 
    \left( C_F \mathcal{R}_\text{sing} + C_A \mathcal{J} \right)
    \,,\, \text{ where } \,
    \mathcal{R}_\text{sing}(\bm{\kappa}, \bm{\kappabar}) \mydef 
    \mathcal{R}(\bm{\kappa}) 
    + \mathcal{R}(\bm{\kappabar}) 
    - 2\mathcal{J}(\bm{\kappa}, \bm{\kappabar}) 
    \,,
\label{eq:emission-rate}
\end{equation}
\noindent with $\mathcal{R}(\bm{\kappa}) = \mathcal{J}(\bm{\kappa},\bm{\kappa})$. As such, this calculation reduces to the interference term $\mathcal{J}$, expressed in terms of the gluon momentum broadening $\mathcal{P}$, the medium-induced emission kernel $\mathcal{K}$, and a decoherence factor $\mathcal{S}_\text{med} =1-\Delta_\text{med}$, written as a function of the antenna opening angle $\bm{\delta n} = (\bm{\kappa} - \bm{\kappabar})/\omega$:

\begin{equation}
\begin{aligned}
	\mathcal{J}(\bm{\kappa},\bm{\kappabar}) 
    =
	\text{Re}	
	\int^\infty_0 \diff t \,
	\int^{\infty}_t \diff\bar{t}
    &
	\int_{\bm{q}_1 \bm{q}_2}
    e^{ i t \left(\tfrac{\omega}{2} \bm{\dn}^2 - \bm{q}_1 \cdot \bm{\dn} \right)}
	\bm{q}_2 \cdot \left(\bm{q}_1 -\omega\bm{\dn} \right)
    \\ 
    &
    \times\,
    \mathcal{P}(\bm{\kappa} - \bm{q}_2; \infty, \bar{t})
	\,\mathcal{K}(\bm{q}_2, \bar{t}; \bm{q}_1, t)
	\,[1-\Delta_{\rm med}(t)]    
    + (\bm{\kappa} \leftrightarrow \bm{\kappabar})
    \,.
\end{aligned}
\label{eq:J-term}
\end{equation}

This formalism requires as modelling choices the medium density $n(t)$ and a momentum transfer rate $V(\bm{q}^2)$, entering into the dipole cross-section $\sigma(\bm{q}) = -V(\bm{q}^2) + (2\pi)^2\delta^2 (\bm{q}) \int_{\bm{\ell}}V(\bm{\ell}^2)$. These quantities determine the evolution of the gluon broadening and emission kernel by means of propagator equations%
\footnote{
$\Delta_\text{med}$ obeys a similar equation, but also admits a closed form: $1-\Delta_\text{med} = \exp\{-\tfrac{1}{2} \int^t_0 \diff\xi n(\xi) \sigma(\xi \bm{\delta n})\}$, where the coordinate space dipole cross-section is computed for realistic rates in \cite{Barata:2020sav}.
} %
for $\mathcal{G} = \{\mathcal{P},\mathcal{K}\}$ with initial conditions $\mathcal{G}_0 = \{\mathcal{P}_0,\mathcal{K}_0\}$:
\begin{align}
    \mathcal{G}(\bm{q}_2, \bar{t}; \bm{q}_1, t)
    &=
    \mathcal{G}_0(\bm{q}_2, \bar{t}; \bm{q}_1, t) 
    -
    \int^{\bar{t}}_{t} \diff s 
    \int_{\bm{u} \bm{v}}
    \mathcal{G}_0(\bm{q}_2, \bar{t}; \bm{v}, s)
    \dfrac{n(s) \sigma(\bm{v} - \bm{u})}{2}
    \mathcal{G}(\bm{u}, s; \bm{q}_1, t)
    \,,
\label{eq:dyson-eqs}
    \\
    \mathcal{K}_0(\bm{q}_2, \bar{t}; \bm{q}_1, t)
    &=
    (2\pi)^2\delta^2(\bm{q}_2 - \bm{q}_1) \, e^{-\im \tfrac{\bm{q}^2_1}{2\omega}(\bar{t} - t)}
    \,,\quad 
    \mathcal{P}_0(\bm{q}_2, \bar{t}; \bm{q}_1, t)
    =
    (2\pi)^2\delta^2(\bm{q}_2 - \bm{q}_1)    
    \,,
\label{eq:dyson-init}
\end{align}

Since including the full momentum range of $V(\bm{q}^2)$ complicates the evaluation of $\mathcal{K}$, typical approaches consider regimes where it has an analytical closed form, such as \textit{multiple soft} and \textit{single hard momentum transfers}. 
The former considers momentum exchanges below a thermal scale $\bm{q}^2 \lesssim \mu^2$, according to the \textit{harmonic oscillator approximation} $n(t)\sigma(\bm{x})\simeq \hat{q}(t)\bm{x}^2/2$, while the latter employs an \textit{opacity expansion} in powers of $\chi = n(t)L$, with $L$ as the medium length. The first order in this expansion is known as the GLV approximation \cite{Gyulassy:2000er}.

As neither approach covers the entire phase-space\footnote{%
A hybrid approach, the \textit{Improved Opacity Expansion} \cite{Barata:2020sav}, has been used to compute this process \cite{Kuzmin:2025fyu}.
}, this work aims to generalise the methods first suggested in \cite{Caron-Huot:2010qjx} and implemented in \cite{Andres:2020vxs} to compute the spectrum of soft medium-induced gluons emitted by a $q\bar{q}$ antenna. 
%
We begin by subtracting the vacuum result $\mathcal{J}^\text{vac}(\bm{\kappa},\bm{\kappabar}) = \tfrac{4\omega^2 \bm{\kappa}\cdot\bm{\kappabar}}{\bm{\kappa}^2 \bm{\kappabar}^2}$, as well as using eqs.~\eqref{eq:dyson-eqs} and \eqref{eq:dyson-init} to reorganise eq.~\eqref{eq:J-term} into:
\begin{equation}    
\begin{aligned}
    &\mathcal{J}(\bm{\kappa},\bm{\kappabar}) - \mathcal{J}^\text{vac}(\bm{\kappa},\bm{\kappabar}) 
    =
    A(\bm{\kappa}, \bm{\kappabar}) 
    + B(\bm{\kappa}, \bm{\kappabar}) 
    + \frac{1}{2} C(\bm{\kappa}, \bm{\kappabar}) 
    + \frac{1}{2} C(\bm{\kappabar}, \bm{\kappa}) 
    ,
    \\
    &A(\bm{\kappa}, \bm{\kappabar}) = 
    \text{Re}\hspace{-0.10cm}
    \int^L_0 \hspace{-0.10cm}\diff s \dfrac{n(s)}{2} 
    \mathcal{S}_\text{med}(s)
    \hspace{-0.10cm}
    \int_{\bm{\ell}\bm{q}}
    \hspace{-0.10cm}
    \mathcal{P}(\bm{\kappa} - \bm{\ell}; L, s) \sigma(\bm{\ell}-\bm{q})
    e^{\im s \tfrac{ \bm{\bar{q}}^2 - \bm{q}^2 }{2\omega}}
    \hspace{-0.05cm}
    \left[
    \mathcal{J}^\text{vac}(\bm{\ell},\bm{\bar{\ell}})
    \hspace{-0.05cm}
    -
    \hspace{-0.05cm}
    \mathcal{J}^\text{vac}(\bm{q},\bm{\bar{q}})
    \right]
    ,
    \\
    &B(\bm{\kappa}, \bm{\kappabar}) = 
    -\int^L_0 \diff s \dfrac{n(s)}{2} 
    \int_{\bm{\ell}\bm{q}}
    \mathcal{P}(\bm{\kappa} - \bm{\ell}; L, s) \sigma(\bm{\ell}-\bm{q})
    \left[
    \mathcal{J}^\text{vac}(\bm{\ell},\bm{\bar{\ell}})
    -
    \mathcal{J}^\text{vac}(\bm{q},\bm{\bar{q}})
    \right]
    ,
    \\
    &C(\bm{\kappa}, \bm{\kappabar}) =
    \text{Re} 
    \frac{4\omega}{\im}
    \int^L_0 \diff s \dfrac{n(s)}{2}
    \int_{\bm{\ell}\bm{q}\bm{u}} 
    \int^{s}_{0} \diff t \,
    \mathcal{S}_\text{med}(t) \,
    e^{-\im t \tfrac{ \bm{\bar{q}}^2 - \bm{q}^2 }{2\omega}} 
    \left(
    \dfrac{\bm{u}}{\bm{u}^2}
    -
    \dfrac{\bm{\ell}}{\bm{\ell}^2}
    \right)\cdot\bm{\bar{q}}
    \\
    &\hspace{0.50\textwidth}
    \times
    \mathcal{P}(\bm{\kappa} - \bm{\ell}; L, s)  \,
    \sigma(\bm{\ell}-\bm{u}) \,
    \mathcal{K}(\bm{u},s;\bm{q},t) 
    \,,
\end{aligned}
\label{eq:J-results}
\end{equation}
%
\noindent where `barred' integration variables denote $\bm{\bar{v}} = \bm{v} - \omega\bm{\delta n}$.

From eqs.~\eqref{eq:J-results} we note how $A$ and $B$ are symmetric under $\bm{\kappa} \leftrightarrow \bm{\kappabar}$; and $A(\bm{\kappa}, \bm{\kappa}) + B(\bm{\kappa}, \bm{\kappa}) = 0 $. All this suggests a separation of $\mathcal{R}_\text{sing}$ in eq.~\eqref{eq:emission-rate} into quark and anti-quark assigned pieces, with the following azimuthal averages:
\begin{align}
    \int \dfrac{\diff\varphi^{}_{\bm{k}}}{2\pi} \mathcal{R}_\text{sing} =
    \underbrace{
    \int \dfrac{\diff\varphi^{}_{\bm{\kappa}}}{2\pi} 
    \Big[
    C(\bm{\kappa},\bm{\kappa}) 
    - C(\bm{\kappa},\bm{\kappabar})
    - A(\bm{\kappa},\bm{\kappabar}) 
    - B(\bm{\kappa},\bm{\kappabar})
    \Big]
    }_\text{Quark Contribution}
    +
    \underbrace{    
    \int \dfrac{\diff\varphi^{}_{\bm{\kappabar}}}{2\pi}
    \, (\bm{\kappa} \leftrightarrow \bm{\kappabar})
    }_\text{Anti-Quark Contribution}
    \,,
\end{align}
\noindent where the left-hand averages over the azimuthal angle of the gluon momentum $\bm{k}$, while those on the right-hand side are relative to  $\bm{\kappa}$ ($\bm{\kappabar}$) i.e., about the axis defined by the $q$ ($\bar{q}$) 3-momentum. %
Since $C$ has the same structure as single-quark emissions, its numerical evaluation has been described in \cite{Andres:2020vxs}. Therefore we focus on $B$ as an example, isolating its integrand:
\begin{equation}
    \psi^{}_{B}(|\bm{\kappa}|=\kappa;\tau,s) \mydef
    -
    \int\frac{\diff\varphi^{}_{\bm{\kappa}}}{2\pi}
    \int_{\bm{\ell}\bm{q}}
    \mathcal{P}(\bm{\kappa} - \bm{\ell}; \tau, s) 
    \dfrac{ \sigma(\bm{\ell}-\bm{q}) }{2}
    \left[
    \mathcal{J}^\text{vac}(\bm{\ell},\bm{\bar{\ell}})
    -
    \mathcal{J}^\text{vac}(\bm{q},\bm{\bar{q}})
    \right]
    \,,
\end{equation}
\noindent which, according to the differential form of eqs.~\eqref{eq:dyson-eqs}~and~\eqref{eq:dyson-init}, obeys the initial value problem
\begin{align}
    \partial_\tau \psi^{}_{B}(\kappa;\tau,s) &=
    - n(\tau) \, 
    \int \dfrac{\diff q \,q}{2\pi} \,
    M_0 (\kappa, q; \mu)
    \left[
    \psi^{}_{B}(\kappa;\tau,s) - \psi^{}_{B}(q;\tau,s)
    \right]
    \,,
\label{eq:psiB-diffeq}
    \\
    \psi^{}_{B}(\kappa;\tau=s) &=
    4\omega^2
    \int \dfrac{\diff q \, q}{2\pi}
    M_0 (\kappa, q; \mu)
    \left[
    \dfrac{\Theta(\kappa-\omega\,\delta n)}{\kappa^2}
    - 
    \dfrac{\Theta(q-\omega\,\delta n)}{q^2}
    \right]
    \,,
\label{eq:psiB-init}
\end{align}
\noindent where $M_0(\kappa,q;\mu) \mydef \int \frac{\diff\phi}{4\pi} V(|\bm{\kappa}-\bm{q}|;\mu)$ with $\phi$ as the angle between $\bm{\kappa}$ and $\bm{q}$.
%

We proceed as follows: evaluate the initial condition for $\psi_B(\tau=s)$ in $(s,\kappa,\omega)$ space and evolve it to $\tau = L$. The $B$ contribution is then $\smallint \diff\varphi^{}_{\bm{\kappa}} B(\bm{\kappa},\bm{\kappabar}) / (2\pi) = \smallint^{L}_{0} \diff s \, n(s)\, \psi_{B}(\kappa;L,s)$, and the same method holds for $A$, albeit with different initial conditions.
%
%

\section{Numerical Results}

This section presents preliminary results for the Yukawa interaction rate $V(\bm{q}^2)=8\pi\mu^2/(\bm{q}^2 + \mu^2)^2$ and constant medium density $n(t) = n_0 \Theta(L-t)$. The relevant parameters are then given by the opacity $\chi=n_0L$, the characteristic energy $\bar{\omega}_c=\mu^2L/2$, and decoherence angle $\bar{\theta}_c = 1/\mu L$.

\begin{figure}[ht]
\centering
\includegraphics[width=.60\textwidth]{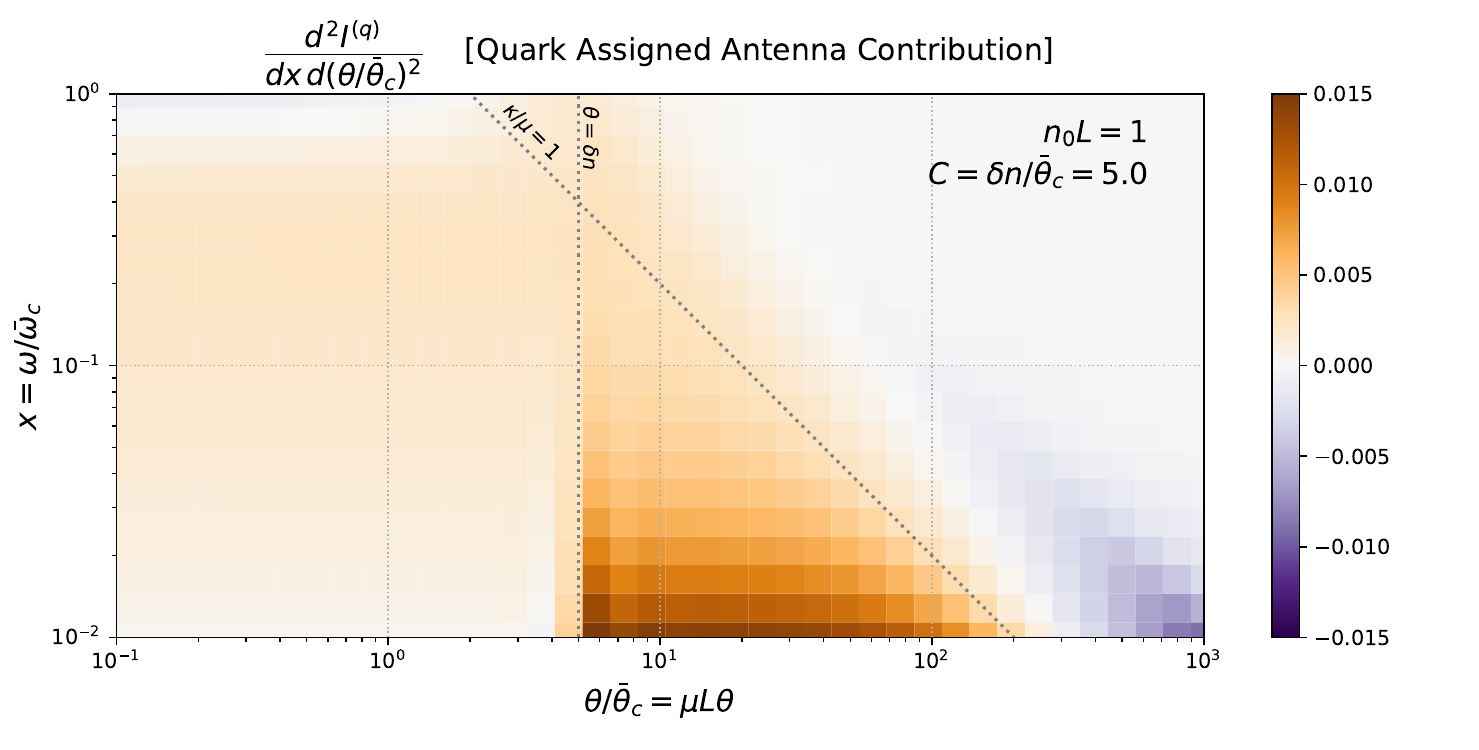}
\includegraphics[width=.35\textwidth]{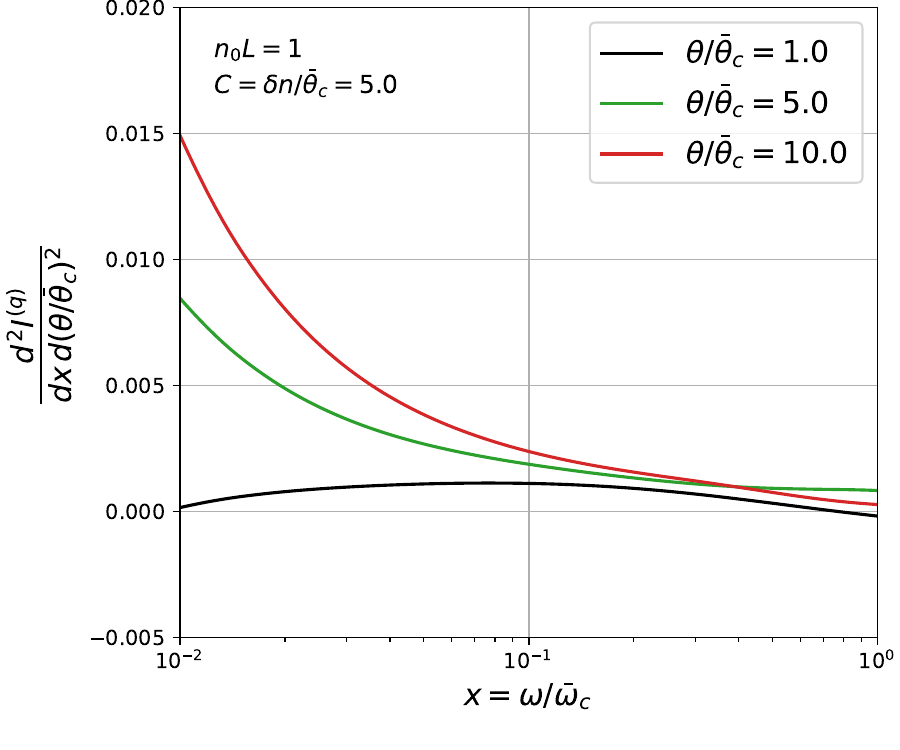}
\caption{
\textbf{(Left)} %
Medium-induced gluon spectrum for a constant density medium with $n_0L = 5$ and a Yukawa scattering rate. %
The horizontal axis represents a rescaled emission angle and the vertical axis a rescaled energy. %
The $q\bar{q}$ antenna opening is $\delta n = 5 \, \bar{\theta}_c$.
\textbf{(Right)} %
Constant angle cuts of the gluon emission spectrum, for $\theta/\bar{\theta}_c = \{1,5, 10\}$ in black, green, and red respectively.}
\label{fig:Lund-Plane-Yukawa}
\end{figure}

Following the formulation in the previous section we focus on the `Quark Assigned Contribution', averaged about the axis defined by the quark's 3-momentum, for which fig.~\ref{fig:Lund-Plane-Yukawa} (left) presents the double differential emission rate in $(\omega/\bar{\omega}_c, \theta/\bar{\theta}_c)$ phase-space, where $\theta = \kappa/\omega$.

Qualitatively, the gluon emission spectrum shows the expected behaviour,
with a constant emission spectrum in the collinear region ($\theta < \delta n$) due to the dominance of the $C(\bm{\kappa}, \bm{\kappabar})$ term. In contrast, out-of-cone radiation ($\theta > \delta n$) is strongly enhanced up to $\kappa \lesssim \mu$, as expected \cite{Mehtar-Tani:2010ebp} from the opening of an anti-angular ordered region of phase-space.
In this anti-angular ordered region soft emissions are particularly enhanced as shown in the right panel of fig.~\ref{fig:Lund-Plane-Yukawa}, where distributions  become more peaked for low $\omega/\bar{\omega}_c$ with increasing values of $\theta/\bar{\theta}_c$.

\section{Summary}

We generalised a method first introduced in \cite{Andres:2020vxs} in the context of gluon radiation from a single quark to the case of coherent radiation from a $q\bar{q}$ pair. By casting the spectrum as the solution to propagator equations, it can be computed numerically for any realistic scattering rate outside of typical approximations.
We presented preliminary results for a Yukawa potential and constant density medium, in the form of the $(\omega, \theta)$ medium-induced gluon spectrum.

In future, these techniques can be applied to the Hard Thermal Loop scattering rate and compared with the harmonic oscillator, GLV, and Improved Opacity Expansion, to better classify their accuracy in different phase-space regions. Further, different observables can be explored by including the quark and anti-quark contributions. 
In the longer term, this work paves the way for precision calculations of jet substructure in heavy ion collisions.

\noindent \textbf{Acknowledgements} AC thanks João Silva, João Barata, and Konrad Tywoniuk for their helpful comments. 
%
%
%
%
This work is supported by: European Research Council project ERC-2018-ADG-835105 YoctoLHC; OE--Portugal, Fundação para a Ciência e a Tecnologia (FCT), under ERC-PT A-Projects ``Unveiling''; Xunta de Galicia (CIGUS Network); the European Union ERDF; the Spanish Research State Agency; and the U.S. Department of Energy.
AC acknowledges FCT support under contract PRT/BD/154190/2022.
\bibliography{bibliography} 
%
%
%
%

\end{document}